\documentclass[aps, prd, twocolumn, preprintnumbers, superscriptaddress, nofootinbib, floatfix]{revtex4-1}

\usepackage{amsfonts}
\usepackage{graphicx}
\usepackage{hyperref}

\begin{document}

\title{Eternal Inflation and the Refined Swampland Conjecture}

\newcommand{\FIRSTAFF}{\affiliation{The Oskar Klein Centre for Cosmoparticle Physics,
	Department of Physics,
	Stockholm University,
	AlbaNova,
	10691 Stockholm,
	Sweden}}
\newcommand{\SECONDAFF}{\affiliation{Department of Physics,
	University at Buffalo, SUNY
	Buffalo,
	NY 14260
	USA}}

\author{William H. Kinney}
\email[Electronic address: ]{whkinney@buffalo.edu}
\FIRSTAFF
\SECONDAFF

\date{\today}

\begin{abstract}
I apply recently proposed ``Swampland'' conjectures to eternal inflation in single-scalar field theories. Eternal inflation is a phase of infinite self-reproduction of a quasi-de Sitter universe which has been argued to be a generic consequence of cosmological inflation. The originally proposed de Sitter swampland conjectures \cite{Obied:2018sgi,Agrawal:2018own} were shown by Matsui and Takahashi \cite{Matsui:2018bsy}, and by Dimopoulos \cite{Dimopoulos:2018upl}, to be generically incompatible with eternal inflation. However, the more recently proposed ``refined'' swampland conjecture \cite{Garg:2018reu,Ooguri:2018wrx} imposes a slightly weaker criterion on the scalar field potential in inflation, and is consistent with the existence of a tachyonic instability. In this paper, I show that eternal inflation is marginally consistent with the refined de Sitter swampland conjecture. Thus, if the refined conjecture is correct, the existence of a landscape-based ``multiverse'' in string theory is not incompatible with a self-consistent ultraviolet completion, with significant consequences for model building in string theory. 
\end{abstract}

\maketitle

\section{Single-field slow-roll inflation and the Swampland-de Sitter Conjectures}
\label{sec:introduction}

Inflation, generally defined, consists of a period of vacuum-dominated accelerating expansion in the early universe, which eventually ceases in a period of reheating and an transition to a radiation-dominated hot Big Bang cosmology \cite{Starobinsky:1980te,Sato:1981ds,Sato:1980yn,Kazanas:1980tx,Guth:1980zm,Linde:1981mu,Albrecht:1982wi}. The vacuum dynamics can be generally modeled using one or more scalar order parameters $\phi$ such that the potential energy of the field dominates over the kinetic energy,
\begin{eqnarray}
\label{eq:prho}
&&\rho = \frac{1}{2} \dot\phi^2 + V\left(\phi\right) \simeq V\left(\phi\right),\cr
&&p = \frac{1}{2} \dot\phi^2 - V\left(\phi\right) \simeq - V\left(\phi\right),
\end{eqnarray}
resulting in an equation of state $p \simeq -\rho$ and quasi-de Sitter expansion,
\begin{equation}
a\left(t\right) \propto e^{H t},
\end{equation}
with 
\begin{equation}
\label{eq:Hubble}
H^2 \simeq \frac{1}{3 M_{\rm P}^2} V\left(\phi\right) \simeq {\rm const.}
\end{equation}
Here $a\left(t\right)$ is the  scale factor, defined in terms of the metric $ds^2$ of a Friedmann-Lemaitr{\'e}-Robertson-Walker (FLRW) spacetime by
\begin{equation}
ds^2 = dt^2 - a^2\left(t\right) d{\bf x}^2,
\end{equation}
and $H$ is the Hubble parameter,
\begin{equation}
H \equiv \frac{\dot a}{a}.
\end{equation}
The equation of motion for a scalar field $\phi$ in an expanding FLRW spacetime is 
\begin{equation}
\label{eq:scalarEOM}
\ddot \phi + 3 H \dot\phi + V'\left(\phi\right) = 0,
\end{equation}
where $V'\left(\phi\right) = \delta V / \delta \phi$ is the variation of the potential with respect to the field. Potential domination requires that the field be slowly rolling, with
\begin{equation}
3 H \dot \phi \simeq - V'\left(\phi\right).
\end{equation}
This condition can be expressed in terms of the {\it slow roll parameter} $\epsilon$, defined as:
\begin{equation}
\epsilon \equiv -\frac{a}{H} \frac{d H}{d a} = \frac{3}{2}\left(1 + \frac{p}{\rho}\right) = \frac{1}{2 M_{\rm P}^2} \left(\frac{\dot \phi}{H}\right)^2.
\end{equation}
In the limit of a slowly rolling field, the slow roll parameter can be approximated in terms of the potential,
\begin{equation}
\label{eq:epsilonSR}
\epsilon \simeq \frac{M_{\rm P}^2}{2} \left(\frac{V'}{V}\right)^2.
\end{equation}
Accelerating expansion requires $\epsilon < 1$, and the limit of de Sitter expansion is $\epsilon \rightarrow 0$. 

In this paper I consider single-field inflation in light of recently proposed swampland conjectures~\cite{Obied:2018sgi,Agrawal:2018own}, which can be stated as inequalities on the potential for the scalar field driving inflation (the {\it inflaton}):
\begin{eqnarray}
\frac{\vert \Delta \phi \vert}{M_{\rm P}} \lesssim \Delta \sim {\cal O}(1)\, \label{eq:sc1}\\
M_{\rm P}\frac{\vert V'\left(\phi\right) \vert}{V} \gtrsim c \sim {\cal O}(1)\,
\label{eq:sc2}
\end{eqnarray}
where $M_{\rm P} \simeq 2.4 \times 10^{18}\,{\rm GeV}$ is the reduced Planck mass, and $c$ is a positive constant of order unity \cite{Obied:2018sgi}. The conjectures state that the inequalities (\ref{eq:sc1},\ref{eq:sc2}) must be satisfied by any low-energy effective field theory (EFT) which has a self-consistent ultraviolet (UV) completion. This first condition does not pose significant difficulty for single-field inflation, and corresponds observationally to a suppressed tensor/scalar ratio via the well-known Lyth Bound for single-field inflation \cite{Lyth:1996im}. The second condition (\ref{eq:sc2}), however, poses more difficulties for single-field inflation, since it is in direct in tension with the slow roll condition (\ref{eq:epsilonSR}). This tension has been explored in a number of recent works, for example Refs. \cite{Kehagias:2018uem,Achucarro:2018vey,Kinney:2018nny,Das:2018hqy,Ashoorioon:2018sqb,Brahma:2018hrd,Lin:2018rnx,Yi:2018dhl,Motaharfar:2018zyb,Lin:2018kjm,Holman:2018inr,Yi:2018dhl}, and applied to eternal inflation in Ref. \cite{Matsui:2018bsy}, where it was shown that eternal inflation is inconsistent with the de Sitter swampland conjectures. More recently, Refs. \cite{Garg:2018reu,Ooguri:2018wrx} proposed a ``refined'' swampland conjecture which is consistent with the existence of tachyonic instabilities, which in the single-field case is
\begin{equation}
\label{eq:SC3}
\left(M_{\rm P}\frac{\vert V' \vert}{V} \gtrsim c\right)\ \mathrm{\bf or}\ \left(M_{\rm P}^2 \frac{V''}{V} \lesssim -c'\right),
\end{equation}
where $c'$ is a second constant of order unity. In this paper, we consider the implications of this slightly weaker constraint on eternal inflation. (See also Ref. \cite{Andriot:2018mav} for a related conjecture, and Refs. \cite{Fukuda:2018haz,Garg:2018zdg,Park:2018fuj,Schimmrigk:2018gch,Cheong:2018udx,Chiang:2018lqx} for discussion of general constraints on inflation models in light of the refined conjectures.)

\section{Eternal Inflation and the Swampland Conjectures}

In addition to the the classical evolution described in Sec. \ref{sec:introduction}, the inflaton will also undergo quantum fluctuations as well, which can be modeled as a Langevin equation by the addition of a stochastic noise term to the classical equation of motion Eq. (\ref{eq:scalarEOM}), 
\begin{equation}
\ddot\phi + 3 H \dot\phi + \frac{d V}{d \phi} = N\left(t,{\bf x}\right),
\end{equation}
where $N\left(t,{\bf x}\right)$ is a Gaussian noise function generated by the quantum fluctuations \cite{Starobinsky:1986fx}. So-called {\it eternal} inflation occurs when quantum fluctuations dominate over classical field evolution, so that the field is as likely to roll {\it up} the potential as downward along the gradient. Therefore, in a statistical sense, inflation never ends: there will always be regions of the universe where the field has fluctuated upward, rather than downward, and inflation becomes a quasi-stationary, infinitely self-reproducing state of eternal inflation \cite{Vilenkin:1983xq,Guth:1985ya,Linde:1986fc,Linde:1986fd,Starobinsky:1986fx}. 
\begin{figure}
\begin{center}
\includegraphics[width=0.45\textwidth]{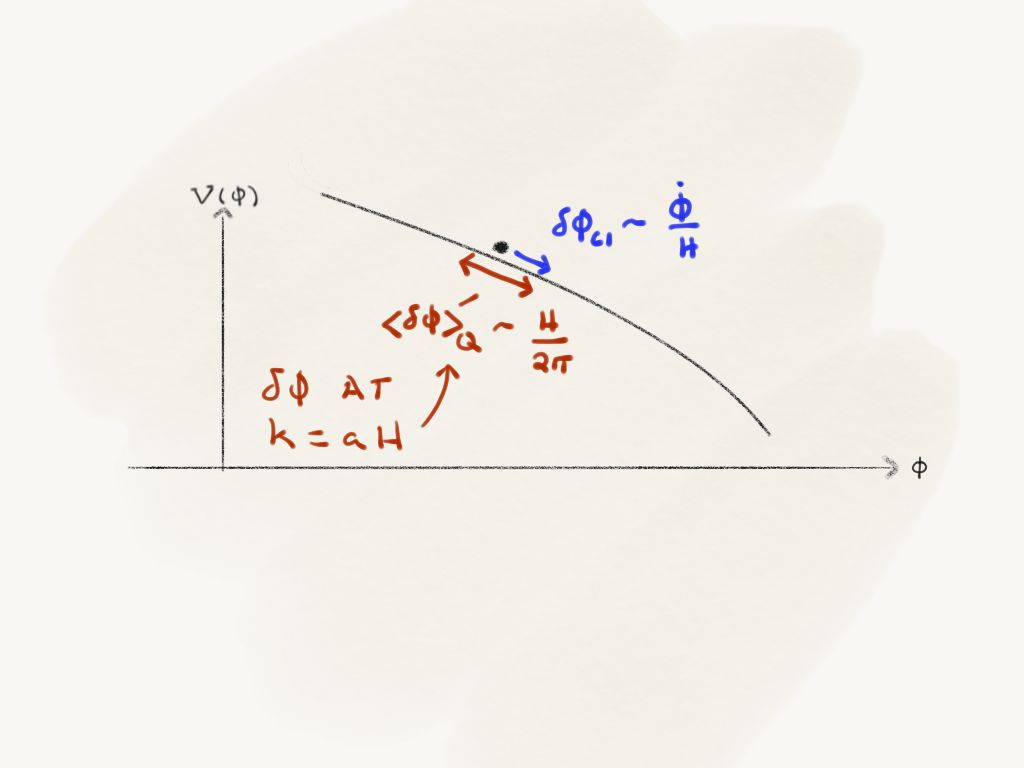}
\end{center}
\caption{Competition between classical evolution $\delta \phi_{Cl}$ and quantum evolution $\left\langle \delta\phi_{Q} \right\rangle$ on a scalar field potential $V\left(\phi\right)$. Eternal inflation occurs when $\left\langle \delta\phi_{Q} \right\rangle > \delta \phi_{Cl}$.}
\label{fig:eternal}
\end{figure}

This competition between classical and quantum evolution can be modeled as a simple inequality (Fig. \ref{fig:eternal}). The amplitude of quantum fluctuations in the inflaton field on scales of order the Hubble length is given by
\begin{equation}
\left\langle\delta\phi\right\rangle_Q \equiv \left\langle \delta \phi^2\right\rangle^{1/2} = \frac{H}{2 \pi}.
\end{equation}
Similarly, the field variation in a Hubble time $t \sim H^{-1}$ is:
\begin{equation}
\delta\phi_{Cl} = \frac{\dot\phi}{H}.
\end{equation}
Therefore, a simple condition for eternal inflation can be written
\begin{equation}
\frac{\left\langle\delta\phi\right\rangle_Q}{\delta\phi_{Cl}} = \frac{H^2}{2 \pi \dot\phi} > 1.\label{eq:eternal}
\end{equation}
It immediately follows that the fraction (\ref{eq:eternal}) is identical to the amplitude of the curvature perturbation for modes crossing the horizon during inflation \cite{Starobinsky:1979ty,Mukhanov:1981xt,Mukhanov:2003xw,Linde:1983gd,Hawking:1982cz,Hawking:1982my,Starobinsky:1982ee,Guth:1982ec,Bardeen:1983qw},
\begin{equation}
P\left(k\right) = \left(\frac{H^2}{2 \pi \dot\phi}\right)^2.
\end{equation}
Accordingly, the condition for eternal inflation is that the amplitude of the comoving curvature perturbation must exceed unity \cite{Goncharov:1987ir,Guth:2007ng},\footnote{While this is a {\it necessary} condition, it is not itself a {\it sufficient} condition for eternal inflation \cite{Vennin:2015hra}.}
\begin{equation}
P\left(k\right) > 1.
\end{equation}
In this limit, we have
\begin{equation}
H^2 / (2 \pi) > \dot\phi,
\end{equation}
so that the slow roll condition is automatically satisfied:
\begin{equation}
\epsilon = \frac{1}{2 M_{\rm P}^2} \left(\frac{\dot \phi}{H}\right)^2 \ll 1.
\end{equation}
This can be stated as the simple condition that eternal inflation {\it requires} nearly de Sitter expansion. It is this requirement that is the source of inconsistency with the originally proposed de Sitter swampland conjectures (\ref{eq:sc1}\ref{eq:sc2}). 

To compare the condition (\ref{eq:eternal}) with the swampland conjectures (\ref{eq:sc1},\ref{eq:sc2}), re-write the condition $P\left(k\right) > 1$  in terms of the slow  roll parameter $\epsilon$ as
\begin{equation} \label{eq:Pk}
P\left(k\right) = \left[\frac{H^2}{8 \pi^2 M_{\rm P}^2 \epsilon}\right]_{k = a H} > 1,
\end{equation}
where $k = a H$ indicates that the quantum fluctuation amplitude is evaluated at horizon crossing. Eternal inflation therefore requires
\begin{equation}
\frac{H}{M_{\rm P}} > 2 \pi \sqrt{2 \epsilon} \simeq 2 \pi M_{\rm P} \left\vert\frac{V'}{V}\right\vert,
\end{equation}
using the slow roll condition (\ref{eq:epsilonSR}). However, to avoid uncontrolled quantum-gravitational corrections, the Hubble parameter $H$ must be small relative to the Planck scale,
\begin{equation}
H \ll M_{\rm P}.
\end{equation}
This results in an inequality, 
\begin{equation}
\label{eq:punchline}
M_{\rm P} \left\vert\frac{V'}{V}\right\vert \ll \frac{1}{2 \pi}.
\end{equation}
This is a general condition for eternal inflation driven by a single canonical scalar field. This condition is immediately seen to be in strong tension with the conjecture (\ref{eq:sc2}),
\begin{equation}
M_{\rm P} \left\vert\frac{V'}{V}\right\vert > c \sim {\mathcal O}\left(1\right).
\end{equation}
This is the result of Matsui and Takahashi \cite{Matsui:2018bsy}. More recently, Dimopoulos considered the possibility of eternal inflation on steep potentials for non-slow roll solutions, with the field reaching a turning point with $\dot\phi = 0$, and concluded that such solutions were likewise incompatible with the de Sitter swampland conjectures \cite{Dimopoulos:2018upl}.

The more recently proposed ``refined'' swampland conjecture weakens this constraint by allowing for quasi-de Sitter expansion driven by a tachyonic instability with $V'/V \simeq 0$, as long as the potential satisfies
\begin{equation}
M_{\rm P} \frac{V''}{V} \leq -c',
\end{equation}
with $c' \sim \mathcal{O}\left(1\right)$. In the case of single-field inflation, this corresponds to ``hilltop''-type models \cite{Kohri:2007gq,Martin:2013tda,Barenboim:2013wra,Coone:2015fha,Huang:2015cke,Vennin:2015egh,Lin:2018rnx,Holman:2018inr}, which were considered in light of the swampland criteria in Refs. \cite{Kinney:2018nny,Lin:2018rnx}. Consistency constraints for eternal inflation in hilltop models were considered in Ref. \cite{Barenboim:2016mmw}: here I summarize the results and apply them to the refined swampland constraint (\ref{eq:SC3}). 

Consider a potential with a tachyonic instability, such that $V'\left(\phi = 0\right) = 0$ and $V''\left(\phi = 0\right) < 0$, which takes the general form in the limit $\phi \rightarrow 0$ of \cite{Kinney:1995cc}
\begin{equation}
\label{eq:V-hilltop}
V\left(\phi\right) = V_0\left[1 - \left(\frac{\phi}{\mu}\right)^2 + \cdots\right].
\end{equation}
We then have
\begin{equation}
\epsilon = \frac{M_{\rm P}^2}{2} \left(\frac{V'}{V}\right)^2 \simeq 2 \left(\frac{M_{\rm P}}{\mu}\right)^2 \left(\frac{\phi}{\mu}\right)^2 \ll 1,
\end{equation}
which is in tension with the swampland conjecture (\ref{eq:sc2}), but is consistent with the refined conjecture (\ref{eq:SC3}). Eternal inflation occurs when $P\left(k\right) > 1$, where
\begin{equation}
P\left(k_{EI}\right) = \frac{\mu^2 H^2}{16 \pi^2 M_{\rm P}^4} \left(\frac{\mu}{\phi_{EI}}\right)^2 = 1,
\end{equation}
so that eternal inflation occurs for a field range $\phi < \Delta \sim \phi_{EI}$, given by
\begin{equation}
\frac{\Delta}{M_{\rm P}} = \frac{1}{4 \pi} \left(\frac{\mu^2 H}{M_{\rm P}^3}\right) \ll 1.
\end{equation}
This is consistent with the swampland conjecture (\ref{eq:sc1}) as long as $H \ll M_{\rm P}$. From Eqs. (\ref{eq:Hubble},\ref{eq:V-hilltop}) with $\phi \ll \mu$, 
\begin{equation}
H^2 \simeq \frac{V_0}{3 M_{\rm P}^2},
\end{equation}
so that
\begin{equation}
\label{eq:hteternal}
\frac{\Delta}{M_{\rm P}} = \frac{1}{4 \sqrt{3} \pi} \left(\frac{\mu^2 \sqrt{V_0}}{M_{\rm P}^4}\right).
\end{equation}
However, the existence of the tachyonic instability in the hilltop case introduces an additional condition for eternal inflation. The expectation value for quantum fluctuations of the field is just the expansion rate,
\begin{equation} \label{dphiQ}
\left\langle \delta\phi\right\rangle_Q = \frac{H}{2 \pi} = \frac{\sqrt{V_0}}{2 \pi \sqrt{3} M_{\rm P}}.
\end{equation}
Due to the presence of a tachyonic instability, the quantum-dominated inflation phase then has finite lifetime. Approximating the field fluctuations as a random walk, the average lifetime $\left\langle t\right\rangle$ of the inflating state will be given by \cite{Starobinsky:1986fx}
\begin{equation}
\sqrt{H \left\langle t\right\rangle} = \frac{\Delta}{\left\langle \delta\phi\right\rangle_Q},
\end{equation} 
or
\begin{equation}
\left\langle t\right\rangle \simeq \frac{4 \pi^2 \Delta^2}{H^3}.
\end{equation}
The probability of inflation ending after time $t$ can then be estimated as \cite{Barenboim:2016mmw}
\begin{equation}
\Gamma_I\left(t\right) \propto e^{-t / \left\langle t\right\rangle} = \exp \left[- \left( \frac{\langle \delta \phi \rangle_Q}{\Delta} \right)^2 H t \right].
\end{equation}
However, during time $t$, an initial Hubble patch of initial volume $\mathcal V\left(0\right)$ will increase in volume  exponentially,
\begin{equation}
\Gamma_{\mathcal V} = \frac{{\mathcal V}\left(t\right)}{{\mathcal V}\left(0\right)} = e^{3 H t},
\end{equation}
so the number of Hubble patches undergoing inflation at time $t$ is
\begin{equation}
\Gamma\left(t\right) \propto \exp{\left[\left(3 - \left( \frac{\langle \delta \phi \rangle_Q}{\Delta} \right)^2 \right) H t\right]}.
\end{equation}
If the exponent is positive, spacetime expansion dominates over the quantum fluctuations. If the exponent is negative, inflation is exponentially quenched. We then have a condition for the consistency of eternal inflation on the hilltop such that expansion dominates over instability,
\begin{equation} \label{eq:EIcondition}
\frac{\Delta}{\langle \delta \phi \rangle_Q} > \frac{1}{\sqrt{3}},
\end{equation}
or from Eq.~$(\ref{dphiQ})$, 
\begin{equation}
\Delta > \frac{\sqrt{V_0}}{6 \pi M_{\rm P}}.
\end{equation}
Applying this to the condition (\ref{eq:hteternal}) for eternal inflation on the hilltop results in the constraint
\begin{equation}
\left(\frac{\mu}{M_{\rm P}}\right)^2 > \frac{2}{\sqrt{3}},
\end{equation}
or, from Eq. (\ref{eq:V-hilltop}),
\begin{equation}
M_{\rm P}^2 \frac{V''}{V} > - \sqrt{3}.
\end{equation}
Thus we see that a period of eternal inflation near the maximum of a potential with a tachyonic instability is in general at least marginally consistent with the refined de Sitter swampland conjecture (\ref{eq:SC3}). This is the main result of this paper.

Note that this is equivalent to a constraint on the second slow roll parameter
\begin{equation}
\eta \equiv M_{\rm P}^2 \frac{V''}{V}.
\end{equation}
Therefore, models which satisfy the refined swampland conjecture (\ref{eq:SC3}) will automatically be in strong tension with constraints from data \cite{Kinney:2018nny}, since for $\epsilon \ll 1$, the scalar spectral index is given by
\begin{equation}
n_{\rm S} - 1 \simeq 2 \eta \ll 1,
\end{equation}
However, this tension with data is not sufficient to rule out an early period of eternal inflation, since perturbations could be generated by a later phase of (for example) Warm Inflation \cite{Berera:1995ie,Das:2018hqy,Motaharfar:2018zyb}, or inflation involving generation of perturbations by an additional field \cite{Achucarro:2018vey}, while still being consistent with an eternally inflating multiverse.

\section{Conclusions}

I have considered eternal inflation in light of the refined de Sitter swampland conjecture \cite{Garg:2018reu,Ooguri:2018wrx}, which states that a scalar field potential associated with a self-consistent UV-complete effective field theory must satisfy one of the two conditions
\begin{equation}
\left(M_{\rm P}\frac{\vert V' \vert}{V} \gtrsim c\right)\ \mathrm{\bf or}\ \left(M_{\rm P}^2 \frac{V''}{V} \lesssim -c'\right),
\end{equation}
where $c$ and $c'$ are constants of order unity. This is a slightly weaker condition than the original proposal \cite{Obied:2018sgi,Agrawal:2018own}, which was shown by Matsui and Takahashi to be generically incompatible with eternal inflation in slow roll \cite{Matsui:2018bsy}, and by Dimopoulos in the case of non-slow roll solutions \cite{Dimopoulos:2018upl}. In particular, the refined conjecture permits inflation in hilltop-type models, characterized by a tachyonic instability. General conditions for the existence of a phase of eternal inflation in hilltop inflation models were derived by Barenboim, Park, and WHK in Ref. \cite{Barenboim:2016mmw}, which results in a bound on the second derivative of the potential of
\begin{equation}
M_{\rm P}^2 \frac{V''}{V} > - \sqrt{3}.
\end{equation}
Therefore, eternal inflation is marginally consistent with the refined de Sitter swampland conjecture: If the conjecture (\ref{eq:SC3}) holds, an eternally inflating multiverse lies in the string landscape, not the swampland, with significant consequences for model building in string theory \cite{Garg:2018zdg,Dvali:2018jhn}. 

\begin{acknowledgments}
WHK is supported by the Vetenskapsr\r{a}det (Swedish Research Council) through contract No. 638-2013-8993 and the Oskar Klein Centre for Cosmoparticle Physics, and by the U.S. National Science Foundation under grant NSF-PHY-1719690.
\end{acknowledgments}

\bibliography{paper}

\end{document}